\documentclass[twocolumn]{article}
\usepackage{cite}
\usepackage{authblk}
\usepackage{amsmath,amssymb,amsfonts}
\usepackage{algorithmic}
\usepackage{graphicx}
\usepackage{textcomp}
\usepackage{url}
\usepackage{array}
\usepackage{multirow}
\usepackage{longtable}
\usepackage{soul}
\usepackage{hyperref}
\usepackage[toc]{appendix}

\def\BibTeX{{\rm B\kern-.05em{\sc i\kern-.025em b}\kern-.08em
    T\kern-.1667em\lower.7ex\hbox{E}\kern-.125emX}}

\begin{document}

\title{BPFroid: Robust Real Time Android Malware Detection Framework}
\author{Yaniv Agman}
\author{Danny Hendler}

\affil{Department of Computer Science, Ben-Gurion University of The Negev, Beer Sheva, Israel \\
agman@post.bgu.ac.il, hendlerd@bgu.ac.il}

\maketitle

\begin{abstract}
We present BPFroid---a novel dynamic analysis framework for Android that uses the eBPF technology of the Linux kernel to continuously monitor events of user applications running on a real device. The monitored events are collected from different components of the Android software stack: internal kernel functions, system calls, native library functions, and the Java API framework. As BPFroid hooks these events in the kernel, a malware is unable to trivially bypass monitoring. Moreover, using eBPF doesn't require any change to the Android system or the monitored applications. We also present an analytical comparison of BPFroid to other malware detection methods and demonstrate its usage by developing novel signatures to detect suspicious behavior that are based on it. These signatures are then evaluated using real apps. We also demonstrate how BPFroid can be used to capture forensic artifacts for further investigation. Our results show that BPFroid successfully alerts in real time when a suspicious behavioral signature is detected, without incurring a significant runtime performance overhead. \\
\end{abstract}

\providecommand{\keywords}[1]{\textbf{\textit{Keywords:}} #1}

\section{Introduction}
\label{sec:introduction}
Android users are exposed to an increasing number of malicious applications despite the great efforts being made to minimize this exposure both by Google and by security products for the Android OS. Many of the existing security solutions for Android use static analysis methods to detect malicious applications. Although these methods may detect some of them, advanced malware often uses techniques such as obfuscation, java reflection, and dynamic code loading \cite{rastogi2013droidchameleon} for avoiding static detection. Overcoming sophisticated evasion techniques such as these requires using dynamic analysis. Efficiently implementing effective dynamic analysis is not an easy task, however. Since benign applications may use the aforementioned techniques as well (e.g., for preventing reverse engineering), differentiating between malicious and benign applications by dynamic analysis is also difficult. Moreover, malicious applications are able to use anti-analysis techniques to avoid being detected by dynamic analysis \cite{maier2014divide, tam2017evolution}.

Previous works suggest methods to perform dynamic analysis to detect malicious applications in Android \cite{backes2017artist, costamagna2016artdroid, druffel2020davinci, Frida:online, ruan2017analyzing, tam2015copperdroid, xposed:online, yan2012droidscope}, and yet, there are no commonly used security products which practically use these methods. Among the reasons for this, is that only a small portion of these works actually use methods that are applicable to real devices, and even a smaller portion can be regarded as robust. Moreover, some of the proposed solutions are bypassable by a capable malware that is aware of them and some of them are unable to identify the malicious intents of an application as they do not directly monitor Android framework API calls, but only lower level system calls.

According to \cite{qamar2019mobile, yan2018survey}, effective and efficient dynamic malware detection approaches are required. Moreover, previously proposed detection approaches are unable to efficiently monitor running applications and provide effective real-time detection \cite{yan2018survey}.

Analyzing malicious applications inside a sandbox becomes insufficient as attackers respond by techniques that evade such analysis. Malicious applications are often able to detect if they are running in a dynamic sandbox environment by looking for known fingerprints of the environment they are running in \cite{maier2014divide}. Such applications become false negatives of the sandbox detection mechanism and are then allowed to run on real Android devices and apply their malicious payload. A second problem with using sandboxes is that they often work only on specially-crafted environments, and require system resources which are unacceptable on a real device. Another problem is that not all applications or even all executions of an application can be tested in a sandbox, because the application may attempt to evade detection by running its malicious payload only after a specific event or trigger happens (e.g. reboot). For this reason, 0-day attacks, which often use such evasion techniques, may not be detected in sandboxes. In contrast, having an Endpoint Detection and Response (EDR) module that always runs in the background may prevent the above problems.

In this work, we attempt to improve security in Android devices by introducing BPFroid, a novel framework that monitors the behavior of Android applications. As Android is built upon the Linux kernel, we make use of a relatively new Linux technology called eBPF, which is available on recent Android phones by default \cite{androidbpfwebsite}. With eBPF, it is possible to write and compile code in userspace, which can then be injected to a running kernel as a callback to selected triggers. \textit{To the best of our knowledge, BPFroid is the first Android malware detection framework to use eBPF}.

In Android, the functionality available to an application is limited and in order for it to compromise security, it must communicate with the Android system. There are several ways for an application to do that. First, it can use the Android framework API, which is the common way for an application to ask for services from the system. However, this is not the only method---an application can also directly call native library functions, or even use Linux system calls, to which eventually every request from the higher levels will be translated. For this reason, BPFroid traces all of these layers: API method invocations, native libraries function calls, direct system calls and even internal kernel function calls.
BPFroid leverages the fact that the Android framework is generally compiled under the ART runtime \cite{configuringARTwebsite}. While its predecessor, Dalvik, used Just In Time (JIT) compilation, ART uses Ahead Of Time (AOT) compilation for its framework and applications in order to gain better performance in runtime. This allows us to inspect the compiled code and calculate the exact addresses of the Android framework API locations in the memory layout. These addresses are then used by BPFroid to hook to their respective functions.

We perform analytical evaluation of BPFroid by comparing it to other Android malware detection frameworks. We show that BPFroid has advantages over alternative frameworks in almost all aspects. Moreover, since Linux' BPF technology continues to evolve and additional features are added to it over time, the added value provided by BPFroid is likely to increase as well. We demonstrate the potential of BPFroid for enhancing the security of Android devices by developing novel signatures that are able to detect specific suspicious behaviors and capture forensic artifacts. BPFroid is able to detect these signatures in real time on a real device and alert on them.

\vspace{2mm}
\textbf{Our contributions.} To summarize, this work makes the following contributions:
\begin{enumerate}
  \item \emph{BPF-based monitor:} We design and implement a novel approach, called BPFroid, for real time application monitoring, based on the Linux BPF technology. BPFroid traces events of Android framework API calls, native libraries functions calls, system calls, and internal kernel functions. To the best of our knowledge, this is the first Android security framework that is able to trace all of these context levels at once.
  \item \emph{Robust malware detection:} In contrast to other solutions, our method leverages a Linux kernel tracing mechanism without tampering with the system image, the application package, or the kernel. As the hooking mechanism runs inside a virtual machine in the kernel, our approach is robust and can safely run on real devices.
  \item \emph{eBPF based dynamic behavioral signatures:} We demonstrate BPFroid usage by developing novel signatures to detect suspicious behavior and capture forensic artifacts.
  \item \emph{Publication of results:} We released BPFroid and the dynamic behavioral signatures as an open-source project to allow researchers to utilize and extend the results of this work at {\url{https://github.com/yanivagman/tracee/tree/BPFroid}}
\end{enumerate}

\vspace{2mm}
\textbf{Outline.} The rest of this article is organized as follows. In Section \ref{background}, we present the relevant background about eBPF, Linux probes, Android ART and Zygote. We describe the requirements for a malware detection framework in Section \ref{requirements}. We then present the architecture of BPFroid in Section \ref{architecture}. In Section \ref{eval}, we evaluate our system by comparing it to other solutions, testing its performance overhead, and presenting possible use cases. In Section \ref{limitations}, we discuss limitations and possible future extensions of our solution. We conclude the paper in Section \ref{conclusions}.

\section{Background} \label{background}
\subsection{BPF}
BPF \cite{bpf} is a VM in the Linux kernel, allowing a privileged user to load and run bytecode safely in the kernel in response to some chosen events. Since version 3.18 of the Linux kernel, the BPF VM has been extended and was named eBPF (extended BPF). We will use the terms BPF and eBPF interchangeably, as eBPF is the only version currently used in the Linux kernel. In order to trigger a BPF program, one needs to attach it to one or more probes, for example, a kprobe \cite{kprobes} or a uprobe \cite{uprobes}.

\begin{figure}[t!]
  \centering
  \includegraphics[width=0.99\columnwidth]{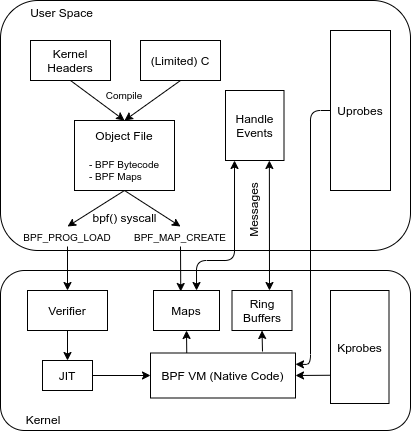}
  \caption{BPF Components.}
    \label{fig:bpf_arch}
\end{figure}

By Just-In-Time (JIT) compilation to native code, BPF offers high performance while executing dynamically loaded eBPF programs. A high level diagram of the BPF components is given in \hyperref[fig:bpf_arch]{Fig. \ref{fig:bpf_arch}}.

\subsection{Kprobes and Uprobes}
Kprobes (kernel probes) and uprobes (user probes) \cite{kprobes, uprobes} are mechanisms in the Linux kernel to dynamically set breakpoints at almost any desired code address, whether in kernel or in user space, specifying a handler routine to be invoked when the breakpoint is hit.

\subsection{ART Runtime}
\begin{figure}[t!]
  \centering
  \includegraphics[width=0.99\columnwidth]{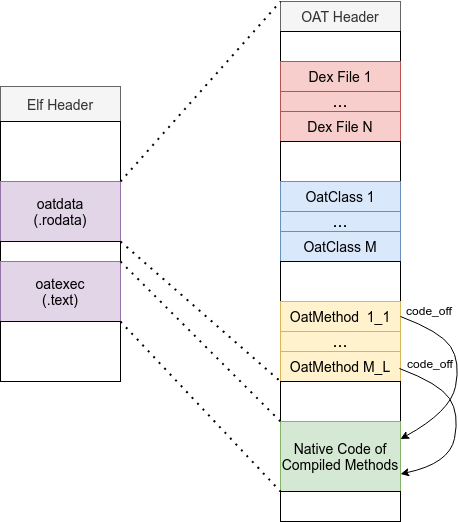}
  \caption{Oat File Format.}
    \label{fig:oat_format}
\end{figure}

Starting from version 5.0, the default Android runtime is ART, which replaced the Dalvik runtime. While Dalvik JIT compiles framework code and applications on demand, ART uses a different approach that performs on-device Ahead-Of-Time (AOT) compilation to transform dalvik bytecode to platform specific native code during installation.
Using AOT compilation has the advantage of achieving better performance at the price of having longer installation times and larger storage requirements. As these disadvantages are small on today's hardware, it is clear why ART was preferred as the default runtime.
After code has been compiled, it is saved on the device in a special file format named oat. This file is actually an ELF executable, which wraps the oatdata section that contains headers and dex files and the oatexec section that contains the compiled code, as can be seen in \hyperref[fig:bpf_arch]{Fig. \ref{fig:oat_format}}.

The oat file format changes between Android versions and there is no official documentation that tracks those changes. Yet, it is possible to reverse engineer the exact format for a given file by looking at the Android Open Source Project (AOSP) \cite{AndroidO76:online} as done by \cite{runtimeoathplatformartGitatGoogle-2020-09-06, sabanal2015hiding}. In order to inspect oat files in a given Android system, it is possible to use the “oatdump” tool, provided by default in the filesystem.

\subsection{Zygote}
Zygote is the parent process of all Android applications, created by the Android runtime at system start.  Zygote is used as a template for creating new processes in the Android system. It includes the first Virtual Machine (VM) and preloads all the relevant Android framework java classes and native libraries into its address space, thus saving to a newly created process the time required to load these resources. Each new application process is a fork of the Zygote process, thus it is a Linux process, which has all the Android framework classes and native libraries preloaded into its address space.

\begin{figure}[t!]
  \centering
  \includegraphics[width=0.7\columnwidth]{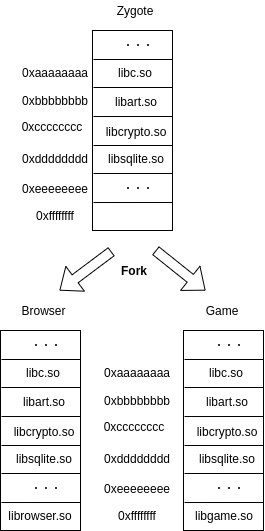}
  \caption{Zygote Fork.}
    \label{fig:zygote}
\end{figure}

By using this approach, the launching time of Android applications is dramatically reduced, because the process address space is already prepared and all that is required is to load the specific application package (APK) into the VM (\hyperref[fig:zygote]{Fig. \ref{fig:zygote}}).

\section{Requirements Analysis} \label{requirements}
In this work, we propose a framework that is able to run on a real device and detect potential malware. In order to do that, we first analyze the requirements for such a framework. To be able to ship the framework as part of an Android system, it should meet the following requirements:

\begin{itemize}
  \item[$R1$] \textbf{Strong security boundary.} The solution should not reside on the same permission level as that of the analyzed application, because otherwise the application can detect it and tamper with the analysis or even disable it.

  \item[$R2$] \textbf{Monitor all system interactions.} To reconstruct the behavioral semantics of the application, the solution should be able to monitor all the interactions of the application with the Android OS. This includes Android framework API calls, native library function calls, and system calls.

  \item[$R3$] \textbf{No system image modification.} The solution should not modify the Android system image or the Linux kernel code.

  \item[$R4$] \textbf{No application modification.} The solution should not change the application package, binaries, or bytecode as these changes either break the same origin policy signature check or can be detected by an app that performs integrity checks.

  \item[$R5$] \textbf{Platform independent.} The solution should be platform independent and should be able to run on real devices.

  \item[$R6$] \textbf{Portable.} Porting the solution to new Android versions should be easy. As new APIs, functions and system calls are added to new Android versions, the solution should make it easy to add or remove tracepoints from it.

  \item[$R7$] \textbf{Scalable.} The solution should be able to monitor multiple applications at once and alert if a malicious behavior is detected in any of them.

  \item[$R8$] \textbf{Safe.} The solution should not affect the system or applications stability. In particular, a software bug in the solution must not crash the target application or the system.

  \item[$R9$] \textbf{Stealth.} The solution should not be easily detectable, because the application can then act benignly and evade detection by the solution.

  \item[$R10$] \textbf{Low overhead.} Because it runs on real devices, the solution should only consume a small portion of the system's resources
\end{itemize}

\section{BPFroid Architecture} \label{architecture}
\subsection{Using BPF for malware detection}
We maintain that BPF is a natural fit for a malware detection system, for several reasons:
\begin{itemize}
  \item Hooking is done in the kernel, thus providing a strong security boundary.
  \item As BPF code gets verified before being inserted to the kernel, and then runs in a VM, programming errors which may cause kernel crashes, hangs, or instability are avoided.
  \item BPF programs can be triggered by probes in user or kernel space, allowing a single mechanism to intercept all events.
  \item BPF supports providing context about the events, including arguments, pid, uid, timestamp, and more.
  \item BPF allows to filter events in the kernel, saving the need to send and parse irrelevant events in userspace.
  \item With BPF it is possible to read and write data to user space memory, hence arguments can be read and even changed.
  \item As code gets to run in kernel space, and compiled to native code, overhead is relatively small.
  \item BPF is maintained as part of the Linux kernel, and new features are constantly being added to it.
\end{itemize}

Because of the aforementioned reasons, we chose to build our malware detection framework on top of BPF. In order to do that, we first had to create an environment where BPF programs will be able to run on Android. Looking at the Android documentation \cite{androidbpfwebsite}, it seems that Android already has the support for running BPF programs. This support, however, is insufficient for our purposes because the Android BPF API seems to be missing some BPF features (e.g. perf buffers for events). For this reason, we chose to take a different approach, where we statically compile an arm64 binary that directly invokes bpf system calls for running our BPF programs, rather than using the API provided by Android to run them.

BPFroid requires the Android kernel to be configured with support for BPF, kprobes and uprobes. While this becomes the standard on new devices (e.g. Google Pixel 4), the device we used for the evaluation of BPFroid, Google Pixel 3a XL, doesn't have all of these requirements, so we had to compile our own kernel image for BPFroid to work on it.

\subsection{Leveraging Zygote and ART}
In addition to using BPF, we take advantage of the following characteristics of Android and ART, which allow us to correctly find the addresses and set uprobes for Android framework API and library functions:

\begin{itemize}
  \item As an Android application is a fork of the Zygote process, \textbf{the memory addresses of the Android framework and native libraries are the same for all Android applications}.
  \item Android ART is set to speed configuration \cite{configuringARTwebsite} when compiling the boot image. This means that \textbf{the Android framework is compiled ahead of time} into oat files.
  \item \textbf{Oat files have a known format} for a given Android version, allowing the extraction of compiled classes and method names, in addition to methods signatures.
  \item \textbf{The Oatdump tool exists on every Android device} and is compatible with the oat version used by that device. As the oat file format may change between Android versions and is not always backward compatible, we use this internal tool available in Android, rather than building our own parser to parse oat files.
\end{itemize}

\subsection{Design Overview}
\begin{figure}[t!]
  \centering
  \includegraphics[width=0.99\columnwidth]{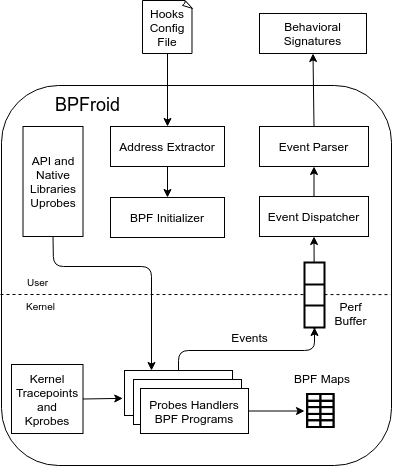}
  \caption{BPFroid Block Diagram.}
    \label{fig:block_diagram}
\end{figure}

A block diagram of BPFroid is given in \hyperref[fig:block_diagram]{Fig. \ref{fig:block_diagram}}.

\begin{figure}[t!]
  \centering
  \includegraphics[width=0.99\columnwidth]{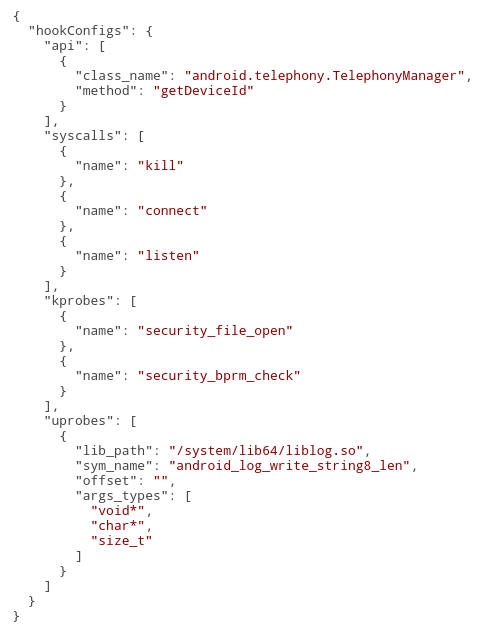}
  \caption{Hooks Configuration File.}
    \label{fig:hooks_json}
\end{figure}

The \textbf{Hooks Configuration File} is where the user of BPFroid defines the events to be traced. There are four types of events which can be traced using BPFroid:
\begin{enumerate}
  \item Android API calls
  \item Linux system calls
  \item Kprobes (used to trace Linux kernel functions)
  \item Uprobes (used to trace Android user space libraries)
\end{enumerate}
An example of a simple hooks configuration file is given in \hyperref[fig:hooks_json]{Fig. \ref{fig:hooks_json}}.

The \textbf{Address Extractor} is responsible for getting and calculating the addresses to attach to the requested events. The process that is used to perform the extraction of addresses is described in the next section.

The \textbf{BPF Initializer} loads the BPF programs into the Linux kernel, attaches the syscalls tracepoints, kprobes and uprobes to their respective handlers, and creates the perf buffer and maps.

The \textbf{Probes Handlers} are BPF programs that are loaded into the Linux kernel. For each type of event (Android API calls, Linux System Calls, Kprobes and Uprobes) there exists a probe handler which is responsible for saving the event, its context, and the provided arguments (whenever possible).

The \textbf{Event Dispatcher} waits for events to be received in the perf buffer and dispatches them to a relevant parser. An event can be any of the events given in the hooks configuration file.

The \textbf{Event Parser} parses the events according to the event context, formatting them in a meaningful way. 

\textbf{Behavioral Signatures} can then apply their rules to detect and alert when a potential threat is found.

\subsection{Implementation}
BPFroid is built upon Tracee \cite{aquasecu65:online}, an open source project developed and maintained by the first author as part of his job at Aqua security \cite{aquasec}. We expanded Tracee's capabilities to support ARM64 devices and added uprobes support so we can trace Android native libraries and Android framework API calls.

BPFroid runs as a privileged application in the system, so it can load BPF programs to the running kernel. As BPF is enabled by default on new Android devices, no kernel nor system image modifications are required. 

To avoid being detected by malware which uses anti-emulation and sandbox detection techniques, we decided to implement BPFroid on a real Android device. By doing so, BPFroid is also able to detect threats in real-time. We evaluate the overhead caused by BPFroid in section \ref{performance}.

\subsubsection{Initialization}
\begin{figure}[t!]
  \centering
  \includegraphics[width=0.99\columnwidth]{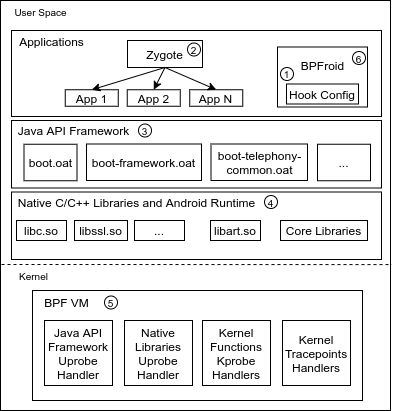}
  \caption{BPFroid Initialization.}
    \label{fig:BPFroid Initialization}
\end{figure}

When BPFroid is started, it performs the following initialization steps, as depicted in \hyperref[fig:BPFroid Initialization]{Fig. \ref{fig:BPFroid Initialization}}.
\begin{enumerate}
  \item \textbf{Obtain the requested events to trace from the hooks configuration file}.
  \item \textbf{Obtain framework API and native libraries paths, along with their respective base addresses in memory.} We first locate the zygote process in the proc filesystem and read its current memory mappings (accessible from /proc/\emph{PID}/maps). We chose to focus our work on 64bit applications. Consequently, we use the zygote64 process (which is the parent of 64bit applications) rather than the zygote process (which is the parent of 32bit applications). We then extract from it the base addresses of all the .oat and .so files, to be used later when we attach the uprobes to the requested functions.
  \item \textbf{Extract oat methods offsets and calculate their exact locations in the apps memory.} We use oatdump with each of the oat files loaded to the zygote address space with execute permission, and search for the requested framework API class and method names. We then extract from the oatdump output the compiled methods offsets and arguments types. Using the offsets and the base addresses extracted in 2), we can calculate the addresses of the compiled functions in the zygote address space and use them to set up uprobes for the requested framework API functions. Finally, we encode the argument types for each of the requested events in a map, setting the key to be the function’s address, and the value to be the encoded arguments types.
  \item \textbf{Extract native libraries function offsets and calculate their exact locations in the applications memory.} For each library function given in the configuration file, we first extract its symbol offset in the .so file. Using this offset and the base addresses extracted in 2), we can calculate the address of the compiled function in the zygote address space and use it to set up a uprobe for the requested library function. We then encode the arguments types given in the config file in a map, setting the key to be the function’s address and the value to be the encoded argument types.
  \item \textbf{Load the BPF programs to the linux kernel and attach them to the probes calculated in 3) and 4) and to the "raw\_syscalls" entry and exit tracepoints.}
  \item \textbf{Wait for events}.
\end{enumerate}

\subsubsection{BPF probe handlers}
The probe handlers are BPF programs which are triggered by one of the traced events. When a handler is triggered, it is first checked that the event needs to be traced according to the filters given to BPFroid. BPFroid does all the filtering in kernel space, so no resources are wasted by the userspace program on irrelevant events.

All of the probe handlers first collect some context about the event. This context is composed of the pid, tid and ppid of the calling thread, a timestamp of the event, the user id, and the name of the process.

As the user of BPFroid can choose any API call or native library function to trace, and as these can change between Android versions, we implemented a generic uprobe handler that handles any of these events. In order to get the event name and metadata (e.g. arguments type), we extract the address of the calling event from the instruction pointer, and use it as a key to a BPF map which was pre-initialized in user space with the events metadata.

The handlers then encode all the collected event data into a message and send it to a ring buffer which is being read from user space.

\subsubsection{Event handling in user space}
BPFroid provides a unified view for all of the probe types. After an event message is read from the ring buffer, it is decoded to get the event id and context. If any additional arguments were provided as part of the event, these are then dispatched to a matching parser according to their type.

For API calls uprobes and library functions uprobes, the instruction pointer address is sent as part of the event message. Using this address as a key, BPFroid can then display the class and method names of an API call, as well as the file and function names of a library function call. At this point, any additional behavioral signatures can test the event and record its data.

\section{Evaluation} \label{eval}
In this section, we evaluate our prototype of BPFroid by first comparing it to other existing techniques that dynamically trace an application. We then measure the performance impact on a real device. Next we present dynamic behavioral signatures which are based on events from BPFroid to alert on suspicious behavior in real time. We then further discuss concrete use cases. For the evaluation of BPFroid, we used a Google Pixel 3a XL running Android 11 and a Linux kernel version 4.9.

\subsection{Analytical Comparison}
Over the years, several techniques were suggested and implemented to perform runtime instrumentation and dynamically trace an application’s behavior. While some of these approaches were shown to have good results for detecting malware, they also suffer from some practical problems.

In the following, we compare existing techniques (AOSP modification, kernel module, application modification, library injection and VMI) by discussing their benefits and shortcomings with respect to requirements R1-R8 which we listed in section \ref{requirements}. \hyperref[table:1]{Table \ref{table:1}} provides a summary of the results of this comparison. A discussion of requirements R9 (stealth) and R10 (low overhead) is deferred to the next subsection.

\begin{table*}[th]
\small
\centering
\caption{Comparison of runtime instrumentation techniques.}
\label{table:1}
\begin{tabular} {|m{1.7cm}|m{1.3cm}|m{1.6cm}|m{1.6cm}|m{1.6cm}|m{1.5cm}|m{1.1cm}|m{1.1cm}|m{1cm}|}
    \hline
     & \textbf{R1.}\newline Strong security boundary & \textbf{R2.}\newline Monitor all \newline system \newline interactions & \textbf{R3.}\newline No system \newline modification \newline & \textbf{R4.}\newline No \newline application \newline modification & \textbf{R5.}\newline Platform \newline independent \newline & \textbf{R6.}\newline Portable \newline \newline& \textbf{R7.}\newline Scalable \newline \newline& \textbf{R8.}\newline Safe \newline \newline\\ 
    \hline
    AOSP\newline Modification & \hfil \includegraphics[width=0.03\textwidth, height=5mm]{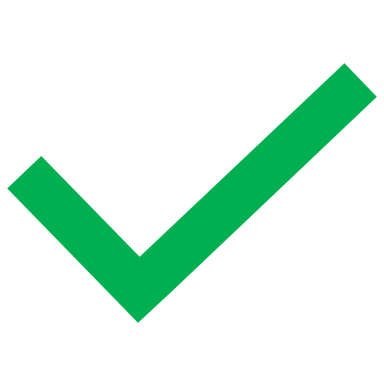} & \hfil \includegraphics[width=0.03\textwidth, height=5mm]{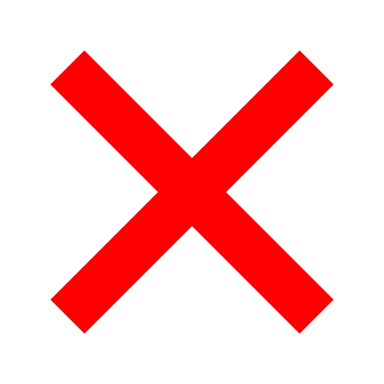} & \hfil \includegraphics[width=0.03\textwidth, height=5mm]{X.png} & \hfil \includegraphics[width=0.03\textwidth, height=5mm]{V.png} & \hfil \includegraphics[width=0.03\textwidth, height=5mm]{V.png} & \hfil \includegraphics[width=0.03\textwidth, height=5mm]{X.png} & \hfil \includegraphics[width=0.03\textwidth, height=5mm]{V.png} & \hfil \includegraphics[width=0.03\textwidth, height=5mm]{X.png}\\ 
    \hline
    Kernel\newline Module & \hfil \includegraphics[width=0.03\textwidth, height=5mm]{V.png} & \hfil \includegraphics[width=0.03\textwidth, height=5mm]{X.png} & \hfil \includegraphics[width=0.03\textwidth, height=5mm]{X.png} & \hfil \includegraphics[width=0.03\textwidth, height=5mm]{V.png} & \hfil \includegraphics[width=0.03\textwidth, height=5mm]{V.png} & \hfil \includegraphics[width=0.03\textwidth, height=5mm]{V.png} & \hfil \includegraphics[width=0.03\textwidth, height=5mm]{V.png} & \hfil \includegraphics[width=0.03\textwidth, height=5mm]{X.png}\\ 
    \hline
    Application\newline Modification & \hfil \includegraphics[width=0.03\textwidth, height=5mm]{X.png} & \hfil \includegraphics[width=0.03\textwidth, height=5mm]{X.png} & \hfil \includegraphics[width=0.03\textwidth, height=5mm]{V.png} & \hfil \includegraphics[width=0.03\textwidth, height=5mm]{X.png} & \hfil \includegraphics[width=0.03\textwidth, height=5mm]{V.png} & \hfil \includegraphics[width=0.03\textwidth, height=5mm]{V.png} & \hfil \includegraphics[width=0.03\textwidth, height=5mm]{X.png} & \hfil \includegraphics[width=0.03\textwidth, height=5mm]{X.png}\\ 
    \hline
    Library\newline Injection & \hfil \includegraphics[width=0.03\textwidth, height=5mm]{X.png} & \hfil \includegraphics[width=0.03\textwidth, height=5mm]{X.png} & \hfil \includegraphics[width=0.03\textwidth, height=5mm]{V.png} & \hfil \includegraphics[width=0.03\textwidth, height=5mm]{V.png} & \hfil \includegraphics[width=0.03\textwidth, height=5mm]{V.png} & \hfil \includegraphics[width=0.03\textwidth, height=5mm]{V.png} & \hfil \includegraphics[width=0.03\textwidth, height=5mm]{V.png} & \hfil \includegraphics[width=0.03\textwidth, height=5mm]{X.png}\\ 
    \hline
    VMI \newline & \hfil \includegraphics[width=0.03\textwidth, height=5mm]{V.png} & \hfil \includegraphics[width=0.03\textwidth, height=5mm]{X.png} & \hfil \includegraphics[width=0.03\textwidth, height=5mm]{V.png} & \hfil \includegraphics[width=0.03\textwidth, height=5mm]{V.png} & \hfil \includegraphics[width=0.03\textwidth, height=5mm]{X.png} & \hfil \includegraphics[width=0.03\textwidth, height=5mm]{V.png} & \hfil \includegraphics[width=0.03\textwidth, height=5mm]{V.png} & \hfil \includegraphics[width=0.03\textwidth, height=5mm]{V.png}\\ 
    \hline
    BPFroid \newline & \hfil \includegraphics[width=0.03\textwidth, height=5mm]{V.png} & \hfil \includegraphics[width=0.03\textwidth, height=5mm]{V.png} & \hfil \includegraphics[width=0.03\textwidth, height=5mm]{V.png} & \hfil \includegraphics[width=0.03\textwidth, height=5mm]{V.png} & \hfil \includegraphics[width=0.03\textwidth, height=5mm]{V.png} & \hfil \includegraphics[width=0.03\textwidth, height=5mm]{V.png} & \hfil \includegraphics[width=0.03\textwidth, height=5mm]{V.png} & \hfil \includegraphics[width=0.03\textwidth, height=5mm]{V.png}\\ 
    \hline
\end{tabular}
\end{table*}

Tracing Android API calls made by an application can be accomplished by \textbf{modifying the Android system code}, as done by Taintdroid \cite{enck2014taintdroid} and Tracedroid \cite{van2013dynamic}. As these changes are part of the Android code, it is possible to trace more than just one application at once (R7: \includegraphics[width=0.02\textwidth, height=3mm]{V.png}) and no modifications are required to the application itself (R4: \includegraphics[width=0.02\textwidth, height=3mm]{V.png}). This solution can theoretically run on real devices (R5: \includegraphics[width=0.02\textwidth, height=3mm]{V.png}); however, as long as the changes are not merged into the AOSP, it is necessary to modify it for each Android version (R6: \includegraphics[width=0.02\textwidth, height=3mm]{X.png}). Other drawbacks of this approach are that it requires system modifications (R3: \includegraphics[width=0.02\textwidth, height=3mm]{X.png}) and that it doesn’t monitor system calls and native library calls as well (R2: \includegraphics[width=0.02\textwidth, height=3mm]{X.png}).

To trace system calls and other internal kernel functions, it is possible to \textbf{modify the Linux kernel}. MADAM \cite{saracino2016madam} and DroidRevealer \cite{ruan2017analyzing}, for example, modify the Linux kernel using a kernel module. Tracing using a kernel module has a strong security boundary (R1: \includegraphics[width=0.02\textwidth, height=3mm]{V.png}), as the kernel has the highest privileges in the Android system. It is also able to trace multiple applications at once (R7: \includegraphics[width=0.02\textwidth, height=3mm]{V.png}), and can run on a real device (R5: \includegraphics[width=0.02\textwidth, height=3mm]{V.png}) as done by DroidRevealer \cite{ruan2017analyzing}. This method, however, has several problems. It requires changing the system image (R3: \includegraphics[width=0.02\textwidth, height=3mm]{X.png}),  so even a single bug in it will crash the whole system (R8: \includegraphics[width=0.02\textwidth, height=3mm]{X.png}). Although using a kernel module can trace system calls, it cannot trace native library functions and Android framework API (R2: \includegraphics[width=0.02\textwidth, height=3mm]{X.png}).

Several \textbf{application rewriting} techniques were proposed for instrumenting an Android application: Modifying the APK, modifying the binary in the oat file, or recompiling the dex files \cite{arzt2017soot, backes2017artist, choi2015api}. These application rewriting techniques can be detected by integrity checks performed by the application (R4: \includegraphics[width=0.02\textwidth, height=3mm]{X.png}). Moreover, as the changes are made to the code of the application itself (R1: \includegraphics[width=0.02\textwidth, height=3mm]{X.png}), it can be evaded as well. Another problem with this kind of instrumentation is that a malicious application can hide its intentions by dynamically loading code (R2: \includegraphics[width=0.02\textwidth, height=3mm]{X.png}). In addition, this solution is not scalable (R7: \includegraphics[width=0.02\textwidth, height=3mm]{X.png}), because it requires patching the application and for different applications, the number of locations that need to be patched to successfully hook a single API may differ.

To avoid modifying the system image (R3: \includegraphics[width=0.02\textwidth, height=3mm]{V.png}) or application itself (R4: \includegraphics[width=0.02\textwidth, height=3mm]{V.png}), it is possible to use \textbf{library injection} techniques as done by frameworks such as Xposed \cite{xposed:online}, Frida \cite{Frida:online}, and ADBI \cite{ADBI:online}, which are used in AppTrace \cite{qiu2015apptrace}, Boxify \cite{backes2015boxify}, MADAM \cite{saracino2016madam} and ARTDroid \cite{costamagna2016artdroid}. With library injection, the tracing logic is injected as a library into the target process address space. As the privilege level of the application and the tracing library are the same (R1: \includegraphics[width=0.02\textwidth, height=3mm]{X.png}), this technique can be detected and bypassed by a competent malware. While it is possible to try and avoid detection by hooking other functions and altering the return values, the defender has no real advantage in this cat and mouse game. Another problem with this approach is that a software bug in the library can crash the application (R8: \includegraphics[width=0.02\textwidth, height=3mm]{X.png}). Moreover, a malware which directly calls a system call or that uses a statically linked native library can bypass the hooking mechanism used by this method.

Yet another approach, which is used by Copperdroid \cite{tam2015copperdroid} and Droidscope \cite{yan2012droidscope}, applies \textbf{Virtual Machine Introspection (VMI)} to reconstruct the OS-level and Java-level semantics of an application by tracking low level VM events, such as system calls. VMI has a strong security boundary (R1: \includegraphics[width=0.02\textwidth, height=3mm]{V.png}) and doesn’t require any system or application modifications (R3: \includegraphics[width=0.02\textwidth, height=3mm]{V.png}, R4: \includegraphics[width=0.02\textwidth, height=3mm]{V.png}). The disadvantage of this approach, however, is that reconstructing the behavioral semantics consumes system resources and may be inaccurate [R2: \includegraphics[width=0.02\textwidth, height=3mm]{X.png}]. As VMI is based on a VM, it is not applicable to real devices (R5: \includegraphics[width=0.02\textwidth, height=3mm]{X.png}) as well. Moreover, running in a simulated environment can be detected by a malware \cite{maier2014divide} and is easy to evade using anti-debugging techniques.

As none of the above techniques can monitor all system interactions at once, some works combined two or more techniques to achieve this. For example, MADAM \cite{saracino2016madam} uses a kernel module to monitor system calls and the Xposed framework to monitor Android API calls.
Using several monitoring techniques, however, can be cumbersome and requires synchronization between the different outputs.

By relying on the BPF technology of the Linux kernel, \textbf{BPFroid} achieves some of the desired requirements “automatically”. A strong security boundary is achieved (R1: \includegraphics[width=0.02\textwidth, height=3mm]{V.png}), as the BPF VM runs in the kernel. In addition, no system or application modifications are required (R3: \includegraphics[width=0.02\textwidth, height=3mm]{V.png}, R4: \includegraphics[width=0.02\textwidth, height=3mm]{V.png}). The BPF programs which instrument the Android system are considered safe (R8: \includegraphics[width=0.02\textwidth, height=3mm]{V.png}) as their code is being validated by the kernel verifier which ensures exactly that. The verifier checks that the programs cannot crash the system, access invalid memory addresses, etc.

BPFroid is able to trace almost any part of the android system (as long as it is natively compiled). It can trace Android framework API, native library functions, system calls and internal kernel functions. By tracing all context levels of an application in a unified manner (R2: \includegraphics[width=0.02\textwidth, height=3mm]{V.png}), BPFroid is able to capture the behavioral semantics of the monitored application.

The only requirement to run BPFroid in an Android system is that the Linux kernel will have BPF support, which already becomes the standard on new Android devices. For this reason, BPFroid is platform independent (R5: \includegraphics[width=0.02\textwidth, height=3mm]{V.png}) and is able to run on real devices. Moreover, porting BPFroid to new Android versions should be easy (R6: \includegraphics[width=0.02\textwidth, height=3mm]{V.png}) as long as the Android framework API code is natively compiled (which is the default for performance reasons).

To monitor a specific application, BPFroid uses the fact that each application in the Android system has a unique user id. To trace a specific application, all that is necessary is to provide BPFroid the application’s UID or package name as an argument to filter by. It is also possible to provide several applications to trace, or even all applications in the system (R7: \includegraphics[width=0.02\textwidth, height=3mm]{V.png}). Moreover, as different users may use the framework in different ways, BPFroid allows the user to choose which tracepoints to activate or filter out.

\subsection{Detectability}
As the BPF VM runs in the Linux kernel, BPFroid has a strong security boundary. This practically means that malware is unable to easily evade it. But can it be easily detected? This is the question we discuss in this section.

Advanced malware that employs anti-debugging techniques may behave benignly in the presence of an analysis environment. Using an emulated environment to perform the analysis can be detected by differentiating it from a real device by looking for device IMEI, cell info, hardware info, and so on \cite{maier2014divide, petsas2014rage}. As BPFroid can run on a real device, this type of detection attempts will fail.

Detecting connected debuggers is done by looking for tracers that use either ptrace or JDWP (to debug the java layer). There are several ways to do that, described in \cite{owasp-mstg:online} and none of them will be able to detect BPFroid, because it does not use ptrace nor jdwp.

Another approach to detect an analysis environment is by looking for suspicious artifacts in the application’s memory or files \cite{owasp-mstg:online}. For example, an injected library can be found in the application’s memory, and it is usually easy to detect that. Other modifications to the application’s code or data in memory can be discovered using integrity checks. The application’s files, which include the manifest, apk, and class (dex) files can also be checked for integrity. BPFroid does not inject any library, nor does it change the application’s memory or files, so this test will fail as well.

Some malware may check if the device it is running on is rooted. As BPFroid requires root permissions to run, this check will succeed. It is possible to avoid rooting by distributing BPFroid as part of the system image; We further discuss this idea in section \ref{use-case}.

So far, we explained why current anti-analysis techniques will fail to discover that BPFroid is deployed on a real device. But can new techniques arise that will be able to discover bpf specific artifacts? We address this question in the following.

It is possible to enumerate the running bpf programs and other bpf artifacts using the bpf() system call. However, using it requires the process to have the CAP\_SYS\_ADMIN capability \cite{bpf2Linu29:online}, which a regular Android process does not possess.

Using kprobes and tracepoints to trace Linux kernel functions, such as system calls entry points, cannot be detected by a regular process. However, uprobes are also used by BPFroid to trace native library functions and Android framework API. The uprobes mechanism works by setting a breakpoint in the address of the library function or oat compiled method loaded into the process address space. An Android application has access to its own address space, so it can try to detect if such a breakpoint is set. However, as we only set uprobes on system libraries, and not on the application code itself, the application cannot perform quick integrity checks. Moreover, as the application does not know in advance on which environment it will run, it cannot calculate the correct offsets in advance. Yet it is possible that an application will try to detect these breakpoints in runtime, and we may see such anti-analysis techniques developed in the future.

We conclude this discussion by noting that BPFroid cannot be detected using current anti-analysis techniques. New approaches to detect BPFroid may be developed in the future. However, detecting an anti-malware analysis environment is beneficial for a malware primarily if it is done in a sandboxed environment. In this case, the malware can become a false negative of the analysis and then be allowed to apply its malicious payload on the real device. Since BPFroid can be deployed as part of the Android system, evading detection on the real device will not assist the malware in attacking the system.

\subsection{Performance} \label{performance}
To evaluate the performance impact of BPFroid we used a standard benchmark application of Android called Geekbench 5, which is distributed through Google Play and models real-world tasks and applications.

Performance tests were conducted on an octa-core Google Pixel 3a XL device, with a stock Android version 11. Since the Pixel 3a XL device is based on Linux kernel version 4.9, which lacks some of the required eBPF features for ARM64 used by BPFroid (e.g. uprobes support), we backported these missing features and created a modified boot image. The applications installed on the device were only the pre-installed applications, Geekbench 5, and the super user manager, Magisk \cite{magisk}. We ran BPFroid as a privileged native executable on the device.

BPFroid allows the user to choose the number and type of events to trace. For our performance evaluation, we have built two sets of events which we estimate represent the typical “load” of events traced by BPFroid. The first set, which we will refer to as “multi-layer”, includes events from the different layers of the Android software stack, and is composed of 50 Android API functions, 4 native functions, 49 system calls, and 3 internal kernel functions (See Appendix \ref{multilayer} for details). The second set, which we will refer to as “all syscalls”, is composed of all of the available Linux system calls in the OS.

To measure the baseline operation of the system, we first run the benchmark tests without launching BPFroid. For each of the sets (“multi-layer” and “all syscalls”) previously defined, we then run BPFroid in three different ways:
\begin{enumerate}
\item No applications are traced.
\item Only the “Geekbench” app is traced.
\item All user applications (UID > 10000) are traced.
\end{enumerate}

Geekbench results are divided into single-core and multi-core scores, where higher scores represent better performance. We took the average of five runs for each of the reported scores. The benchmark results, summarized in \hyperref[table:2]{Table \ref{table:2}}, show that BPFroid incurs an acceptable overhead.

\begin{table}[th]
\footnotesize
\centering
\caption{Geekbench results averaged over 5 runs, higher is better. Percent differences are given in parentheses.}
\label{table:2}
\begin{tabular} {|c|c|c|c|}
    \hline
    Traced App & Traced Events & Single-Core & Multi-Core \\
    \hline
    \multirow{3}{5em}{None} & None & 337 & 1225 \\
    & Multi-Layer & 323 (4.15\%) & 1200 (2.04\%) \\
    & All Syscalls & 323 (4.15\%) & 1200 (2.04\%) \\
    \hline
    \multirow{2}{5em}{Geekbench} & Multi-Layer & 313 (7.12\%) & 1185 (3.27\%) \\
    & All Syscalls & 301 (10.68\%) & 1095 (10.61\%) \\
    \hline
    \multirow{2}{5em}{All} & Multi-Layer & 313 (7.12\%) & 1172 (4.33\%) \\
    & All Syscalls & 301 (10.68\%) & 1078 (12\%) \\
    \hline
\end{tabular}
\end{table}

Looking at the results where BPFroid runs in the background without tracing any application shows the impact of running BPFroid on processes which are not traced. BPFroid uses the raw\_syscalls entry and exit tracepoints, which are triggered for every system call that happens in the system. For every event, BPFroid checks if it was chosen, and if the current process should be traced, which in combination with the use of raw\_syscalls tracepoints explains the minimal penalty of approximately 2\%-4\% we got in the results.

Tracing all system calls clearly requires more resources than tracing the multi-layer set of events since it traces more events, as well as events which are more frequent (e.g. read, write system calls). We believe that many of these frequent system calls are often not required for security purposes, and so can be removed from the traced set of events, thus incurring even lower overhead.

Another point to note, is that there is no change in single core workload overhead between tracing a single application (Geekbench) or all user applications. This can be explained by the fact that BPFroid does not require more than a single cpu to process the events in userspace, so having another single-core workload can be processed by another CPU in parallel without affecting performance. On the other hand, multi-core workload is affected by the number of traced applications because it competes with BPFroid on the same processing power.

\subsection{eBPF Based Dynamic Behavioral Signatures}
Other than the ability to trace the different layers of an Android system, the true power of using eBPF lies in its ability to safely run custom logic in the kernel with full system visibility, as well as the ability to share data with a userspace program through the use of maps and shared buffers. We demonstrate these capabilities by developing two simple, yet powerful, eBPF based behavioral signatures which we then evaluate on our test device using recent malware samples from the years 2019-2020 \cite{sk3ptre, malwares2019, malwares2020} that include 200 samples. We also show how we capture forensic artifacts in real-time for further forensic investigation.

\begin{figure*}[t!]
  \centering
  \includegraphics[width=0.95\textwidth]{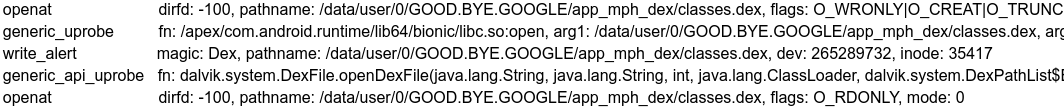}
  \caption{Trimmed output of BPFroid showing a dex file that was dropped by Covid\_Cerberus malware.}
    \label{fig:dropped-dex}
\end{figure*}

\subsubsection{Dropper detection}
A technique that is widely used by malware involves the usage of “droppers”, which are simple applications that will usually look like any other legitimate application and do not request many permissions or perform actions that can be classified as malicious. These droppers can then unpack or download a second stage malicious payload which has more capabilities and is far more potent. This technique allows the distribution of such innocent looking droppers using the Play Store.

To identify such droppers, we hook the vfs\_write(v) kernel functions, which are called whenever a file write is performed in the system. We then look for writes that are made to the first four bytes of some file and check if these bytes are equal to the magic number of either an elf file (“\textbackslash x7fELF”), a dex file (“dex\textbackslash n”), or an archive file (“PK\textbackslash x03\textbackslash x04”). If such a match occurs, we alert the user and provide the pathname of the written file, as well as its device ID and inode number.

Implementing the same detection logic by only tracing system calls wouldn’t be as simple and elegant as tracing the vfs\_write(v) kernel functions, since we would have to track the open file descriptors and the file offset of each of the monitored application processes in order to know to which file the write is being made, and to which offset the data is being written. Tracking this information would require us to monitor several different system calls that are able to change the open file descriptors table or a file’s offset. In contrast, by hooking the vfs\_write(v) kernel functions we can access the internal kernel file struct, which enables us to obtain the file device ID and inode number that uniquely identify the file in the system, as well as to extract the full path of the file. These functions also include as an argument the offset of the file that is being written to, which we then use to directly check if the file’s header was modified.

BPFroid can also be used to capture all of the data that was written to a file in real time. In order to do so, it writes the data which is given as a pointer argument to the vfs\_write(v) functions, as well as some metadata that includes the inode number and the current offset in the file, into a buffer which is shared between the kernel and the user space. This ability can then be leveraged for performing further forensics investigation as we will see next.

\begin{table*}[th]
\footnotesize
\centering
\caption{Dropped Files With Elf Magic Number.}
\label{table:dropped-elf}
\begin{tabular} {|m{4.8cm}|m{10.5cm}|}
    \hline
    APK hash (MD5) & Dropped File(s) \\
    \hline
    e537a5ed354b5cbcce3052901326629b & /data/data/lmh.android.gjbus/curl \\
    \hline
    fef9f84d352d11f2b35b2bb62d15b768 & /data/user/0/com.kevin.quick.app/.jiagu/libjiagu.so,  /data/user/0/com.kevin.quick.app/.jiagu/libjiagu\_64.so \\
    \hline
    0e6352d674165dbbb34808c5f25f0aa4 & /data/user/0/walking.pedometer.fit.stepcounter/app\_bin/daemon \\
    \hline
    19eff67bb340bc6561cbb2c8a84fd460 & /data/user/0/com.eraser.ygycamera.background/app\_daemon/godaemon \\
    \hline
    bdafb22ae3c9c0059f17fc9fb6dabb39 & /data/user/0/com.prosun.beauty.selfie.camera/files/libexec.so \\
    \hline
    d037c03d526c9a130b8dcd346a95d858 & /data/user/0/com.smartapp.donottouch/files/3fab4f41/raw,  /data/user/0/com.smartapp.donottouch/cache/bad72985 \\
    \hline
    35027baae5178b5642dd6fa17857ff51 & /data/user/0/com.hua.ru.quan/files/ali-s/libzuma.so  \\
    \hline
    c907d74ace51cec7cb53b0c8720063e1 & /data/user/0/org.rabbit/files/rssocks \\
    \hline
    ecac763feff38144e2834c43de813216 & /data/user/0/com.udxlbuno.plwnnhop/files/437602.so \\
    \hline
\end{tabular}
\end{table*}

\begin{table*}[th]
\footnotesize
\centering
\caption{Dropped Files With Dex Magic Number.}
\label{table:dropped-dex}
\begin{tabular} {|m{4.8cm}|m{10.5cm}|}
    \hline
    APK hash (MD5) & Dropped File(s) \\
    \hline
    b6656bb8fdfb152f566723112b0fc7c8 & /data/user/0/com.firebear.androil/files/AdDex.4.0.1.dex \\
    \hline
    ac67f1b22d6c7812003609de284a9ad9 & /data/user/0/access.pact.agent/app\_DynamicOptDex/DAOO.json \\
    \hline
    d9945719fa3177fa4e80261a2fb72373 & /data/user/0/ufD.wykyx.vlhvh/files/dex \\
    \hline
    8b64643b2dbe2de2b6b06ac31627b7eb & /data/data/com.selfie.beauty.candy.camera.pro/mix.dex \\
    \hline
    26c078fb212b74f36f51bfb45ecad672 & /data/user/0/samsungupdate.com/files/audience\_network.dex, /data/user/0/samsungupdate.com/app\_working/amazon\_ads.dex, /data/user/0/samsungupdate.com/app\_working/applovin.dex \\
    \hline
    193a1b24952cbe1a4ba1e39cbce25535 & /data/user/0/com.photo.watermark.proeditor/files/audience\_network.dex \\
    1416f440d371dc47d03adae29d3f5e64 & /data/user/0/com.scn.ner.pdfdoc.bright.ing/files/prodexdir/00O000ll111l\_1.dex, /data/user/0/com.scn.ner.pdfdoc.bright.ing/files/prodexdir/00O000ll111l\_0.dex, /data/user/0/com.scn.ner.pdfdoc.bright.ing/files/audience\_network.dex \\
    \hline
    ab6571f8c6ca1ba58ed84319afdc61ab & /data/data/con.xhct945.xhch/app\_mph\_dex/classes.dex \\
    \hline
    35027baae5178b5642dd6fa17857ff51 & /data/user/0/com.hua.ru.quan/files/ali-s/dex2oat/test.dex \\
    \hline
    660159f431b5f8ec8c4fed0298168d1a & /data/user/0/GOOD.BYE.GOOGLE/app\_mph\_dex/classes.dex \\
    \hline
    a39439dbe22a5642daf0d90f8f91c330 & /data/user/0/uwr.oqtlajc.wksayc/app\_dex/lo.dex, \newline /data/user/0/uwr.oqtlajc.wksayc/app\_dex/app.dex \\
    \hline
\end{tabular}
\end{table*}

\begin{table*}[th]
\footnotesize
\centering
\caption{Dropped Files With Archive(apk/jar/zip) Magic Number.}
\label{table:dropped-archive}
\begin{tabular} {|m{4.8cm}|m{10.5cm}|}
    \hline
    APK hash (MD5) & Dropped File(s) \\
    \hline
    2ba22b1922de68634d7729175c29bf55 & /storage/emulated/0/.apk.apk \\
    \hline
    4e1af25f84200c7f63e315fe7ca07a9c & /data/user/0/com.uklildk.ftnqxtietapb/app\_files/saxjqez.jar \\
    \hline
    b6656bb8fdfb152f566723112b0fc7c8 & data/user/0/com.firebear.androil/files/972658728/2086782360.jar \\
    \hline
    dc74daf70afc181471f41fd910a0dec0 & /data/user/0/com.wogdjywtwq.oiofvpzpxyo/app\_files/dkodeh.jar \\
    \hline
    af59407f9534c537f885a9bbdba2964b & /data/user/0/dj.czm.ygq/app\_en/dj.czm.ygq.jar \\
    \hline
    715a094b9da332d3aec8cbdef8ff4b32 & /data/user/0/com.Mobiappsfun.TablaaORG2019.ORGPianoGuitarRobabDigital musicApp/files/.MultiDex/adsdk.zip \\
    \hline
    117e1331306fec02b1ffe6b68d148cc9 & /data/user/0/com.dotgears.flappybird/cache/ads2781488618933041481.jar \\
    \hline
    1250f1296c0e92112d554960d4f99710 & /data/user/0/com.legendapp.qrcode20/app\_product\_odex/product.apk \\
    \hline
    cf7f01274b2a79919b714d2b54831f7e & /data/user/0/com.pmzlpj.miwgrglhzd/app\_files/zgvseohc.jar \\
    \hline
    331578141f84b11077e3222b1d7d7785 & /storage/emulated/0/origin/Camera.apk, \newline /storage/emulated/0/origin/Core.apk, \newline
    /storage/emulated/0/origin/Location.apk, \newline /data/user/0/com.isyjv.klxblnwc/Plugin/com.android.dmp.cm/apk/base-1.apk, \newline /data/user/0/com.isyjv.klxblnwc/Plugin/com.android.dmp.c/apk/base-1.apk, \newline /data/user/0/com.isyjv.klxblnwc/Plugin/com.android.dmp.l/apk/base-1.apk \\
    \hline
    3c5abec5b685809a670dee9b729a9096 & /data/user/0/com.contact.withme.texts/files/poroc \\
    \hline
    5f529573d5d4d067700e981f09c48069 & /storage/emulated/0/abc.apk \\
    \hline
    660159f431b5f8ec8c4fed0298168d1a & /storage/emulated/0/Android sIwI Tester/base.apk \\
    \hline
    7107ac3bccd8db274b21f0e494e3eccc & /data/user/0/com.example.eventbot/app\_dex/4c5826ea665a165fac665135f8d18 876.jar \\
    \hline
    bbe84ba545d652d9e06635a6e89d48b5 & /storage/emulated/0/Aarogya Setu Support/base.apk \\
    \hline
    c8e8080c1365da6dc340edc17d86f674 & /data/user/0/com.hmvoice.friendsms/files/baobutong \\
    \hline
\end{tabular}
\end{table*}

We tested this behavioral signature on our malware samples and were able to detect 33 samples that ``dropped'' files\footnote{We do not know the exact number of our malware samples which are indeed droppers.}, out of which 9 samples dropped ELF files (\hyperref[table:dropped-elf]{Table \ref{table:dropped-elf}}), 11 samples dropped DEX files (\hyperref[table:dropped-dex]{Table \ref{table:dropped-dex}}), and 16 samples dropped archive files which are either .apk, .jar or .zip files (\hyperref[table:dropped-archive]{Table \ref{table:dropped-archive}}). In total, 20 of these dropped files were already known to be malicious by VirusTotal \cite{virustotal}. It is interesting to note that 11 of these dropped files were not recognized by VirusTotal, and we uploaded them to it for the first time. Out of these, 7 files were found to be malicious.

\hyperref[fig:dropped-dex]{Fig. \ref{fig:dropped-dex}} shows a trimmed output of BPFroid during the execution of a malware sample named Covid\_Cerberus (MD5: 66c4513025128719dda018820cc0987e). We can see how BPFroid was able to trace events from all of the different layers while the malware dropped a dex file. A file named classes.dex was created first, for which we can see the system call as well as the native libc function call that was used to open it. We then see a write alert that was triggered as the file was written to disk and a call to the openDexFile Android API that was made after the file was written. We captured the dex file as it was written to disk and then uploaded it to VirusTotal which identified it as a malicious Android trojan.

\subsubsection{Privilege escalation detection}
Over the years there have been several vulnerabilities which were exploited by malware to perform privilege escalation in Android. An example of such an exploit is DirtyCow \cite{dirtycow} which has been used by malwares like FinSpy \cite{finspy}, and Pegasus for Android (Chrysaor) \cite{chrysaor} to gain root privileges on an unrooted device. These attacks, although less common, are powerful attacks that enable the malware to gain full control of the device.

We use the fact that each application in Android has a unique UID that remains constant during its runtime to write a simple rule that checks if an application has escalated its privileges. After a process of an Android application is forked, we use its PID as a key to an eBPF hash map and save its UID as the value. Between every two subsequent system calls we then check if the UID of the process changed. We alert the user in case that such a change happened.

From the malware samples we tested, there was no malware that was able to escalate its privileges and change its UID. This was expected as we used a device with a recent Android version 11 and latest security patches, where no known vulnerabilities should exist. Nonetheless, new zero days are yet to be discovered and this type of behavioral signature is likely to discover such exploits in real-time.

\subsection{Real-time Malware Detection Framework} \label{use-case}
According to \cite{hammad2018large}, the majority of anti-malware products are severely impacted by even trivial code obfuscations, indicating their heavy usage of static analysis techniques. Static techniques are also vulnerable to evasive malware that uses dynamic code loading and java reflection \cite{tam2017evolution}. Recent malware may also incorporate advanced anti-detection techniques to evade detection by dynamic analysis systems \cite{gajrania2020effectiveness, maier2014divide, tam2017evolution}. Key examples of such techniques include anti-emulation techniques, application integrity checks, hiding malicious activity deeper in the code, and logic bombs that are triggered by a specific event (e.g. time-period).
For this reason, monitoring an application’s behavior in real-time is crucial for the detection of intrusions or attacks happening at app runtime. Malware detection by monitoring behaviors, evidence, and features can identify zero-day attacks and detect malware after malware infection \cite{yan2018survey}.

Implementing such a method that can be installed in mobile devices, however, is not an easy task. The solution is required to continuously monitor running applications while incurring only a small overhead and without affecting the stability or integrity of the system or any application running in it. On-device real-time detection methods that are both lightweight and effective still do not exist \cite{yan2018survey}. We believe that BPFroid marks a significant step towards obtaining this goal.

BPFroid can be distributed as a privileged system application, providing configurable hooking capabilities to perform real-time monitoring and behavioral analysis. Requiring only a Linux kernel with eBPF support and low overhead, it is possible to run BPFroid as a background service on a real device, without performing any modifications to the Android system or applications.

Using BPFroid, an anti-malware product can choose if it wants to trace the whole system or specific applications, configure a set of functions to hook, and define dynamic behavioral signatures to use. This configuration can be adaptively changed according to system constraints and scenarios. For example, the whole system can be monitored all the time, looking for behavioral signatures which are known to be malicious with high probability. An alternative approach may be monitoring newly installed applications for a given period of time during which any suspicious behavior will be logged. Other than behavioral signatures, the data collected from BPFroid can be used as input to more advanced algorithms (possibly running on remote servers), which can then apply machine learning methods to improve detection results.

\section{Limitations and future work} \label{limitations}
Several potential limitations of our solution are listed below.
\begin{enumerate}
    \item Loading BPF code into the Linux kernel requires elevated privileges (CAP\_SYS\_ADMIN). For this reason, our framework currently needs to run as root. To avoid rooting the device, BPFroid can be distributed as part of the system image. As BPFroid applications are self-contained, no further system image modifications are required.
    \item As we use uprobes to hook Android framework API functions, it is required that these functions will be precompiled. The move to ART’s AOT compilation was done for performance reasons, and we do not expect this will change in future versions of Android. There are, however, some API functions that are not compiled to native code by default and functions which cannot be optimized, and are therefore JIT compiled by ART. Because we didn’t want to modify the Android system image, BPFroid is unable to trace these functions. Another problem of hooking oat compiled methods is that there are cases where two different functions will point to the same compiled code. This can happen, for example, in empty functions (e.g. android.app.Service.onDestroy(), android.app.Service.onCreate()).
    \item BPFroid does not trace the arguments of Android API functions. Extracting Java classes requires reversing the format of primitives (e.g. String) in memory in order to know how to correctly read and parse it. We leave this issue to future work.
    \item Using a debuggable version of an application (by setting the android:debuggable flag to true in the application manifest) will use a “slow path” where the interpreter will be used instead of the oat precompiled code. When this happens, our framework will not intercept Android API calls events. As this happens mostly in a development environment and not on the end user device, we do not consider this a real problem.
    \item Writing BPF hooks has several limitations due to the fact that BPF code gets verified before being inserted into the kernel. For example, loops are not allowed, and code is limited in size. These limitations prevent us from providing Turing-complete hooks, but for our purposes, we could easily extract the information we needed from the hooks.
\end{enumerate}

In this work, we presented a malware detection framework we developed using BPF hooks. There are several directions that can be investigated in future works:
\begin{enumerate}
    \item As BPF technology evolves, it may be possible to implement policy enforcement when malicious behavior is detected. In recent kernels, it is already possible to override certain kernel functions (e.g. system calls) return values. Another recent addition in kernel 5.7 called LSM BPF \cite{LSMBPFProgramsTheLinuxKerneldocumentation-2020-08-31} introduces policy enforcement with BPF on Linux LSM hooks. These methods may be used to enforce kernel level policies.
    In order to enforce policies with uprobes, it may be possible to use the fact that it is possible to write into user space from BPF hooks. By doing so, it might be possible to change function arguments and return values.
    \item In order to make our framework more “Android native”, it could be implemented using Google’s eBPF API. This will allow vendors to integrate our solution more easily as part of the system image.
    \item Using BPF, it is possible to trace almost any kernel function, as well as to use Linux perf events for tracing, for example, CPU performance counters. This should enable expanding BPFroid to detect hardware-based malware attacks, as we can monitor low level kernel functions that communicate with the hardware and access other statistics saved by the kernel.
    \item Future approaches may combine BPFroid with static analysis as another layer in order to achieve better detection results.
\end{enumerate}

\section{Conclusions} \label{conclusions}
We presented the first Android malware detection framework based on eBPF. The underlying eBPF technology enabled us to build a solution with a strong security boundary, which is capable of running in a safe way and trace application to OS interactions in all context levels. We implemented BPFroid and demonstrated how it can run on a real device and detect malicious behaviors in real time using dynamic behavioral signatures, without   incurring significant runtime performance overhead. We have made the BPFroid source code publicly and freely available.

Our analysis and evaluation indicates that using eBPF for malware detection is a promising direction. Future research can explore new ways to enforce policies using it, building on top of BPFroid and improving its detection capabilities.

\bibliographystyle{IEEEtran}
\bibliography{bpfroid}

\newpage

\appendices

\section{Multi-Layer Events Set}
\label{multilayer}

\begin{table}[thb!]
\footnotesize
\centering
\caption{Native library functions in the multi-Layer events set.}
\label{table:4}
\begin{tabular} {|c|c|}
    \hline
    Library & Function \\
    \hline
    libbinder\_ndk.so & AIBinder\_new \\
    libcamera2ndk.so & ACameraManager\_openCamera \\
    libc.so & open \\
    libdl.so & dlopen \\
    \hline
\end{tabular}
\end{table}

\begin{table}[thb!]
\footnotesize
\centering
\caption{Internal kernel functions in the multi-Layer events set.}
\label{table:6}
\begin{tabular} {|c|c|}
    \hline
    \multicolumn{2}{|c|}{sched\_process\_exit} \\
    \multicolumn{2}{|c|}{vfs\_write} \\
    \multicolumn{2}{|c|}{security\_bprm\_check} \\
    \hline
\end{tabular}
\end{table}

\begin{table}[thb!]
\footnotesize
\centering
\caption{System calls in the multi-Layer events set.}
\label{table:5}
\begin{tabular} {|c|c|}
    \hline
    \multicolumn{2}{|c|}{execve} \\
    \multicolumn{2}{|c|}{execveat} \\
    \multicolumn{2}{|c|}{fork} \\
    \multicolumn{2}{|c|}{vfork} \\
    \multicolumn{2}{|c|}{clone} \\
    \multicolumn{2}{|c|}{open} \\
    \multicolumn{2}{|c|}{openat} \\
    \multicolumn{2}{|c|}{stat} \\
    \multicolumn{2}{|c|}{lstat} \\
    \multicolumn{2}{|c|}{bpf} \\
    \multicolumn{2}{|c|}{perf\_event\_open} \\
    \multicolumn{2}{|c|}{access} \\
    \multicolumn{2}{|c|}{faccessat} \\
    \multicolumn{2}{|c|}{unlink} \\
    \multicolumn{2}{|c|}{unlinkat} \\
    \multicolumn{2}{|c|}{symlink} \\
    \multicolumn{2}{|c|}{symlinkat} \\
    \multicolumn{2}{|c|}{chmod} \\
    \multicolumn{2}{|c|}{fchmod} \\
    \multicolumn{2}{|c|}{fchmodat} \\
    \multicolumn{2}{|c|}{chown} \\
    \multicolumn{2}{|c|}{fchown} \\
    \multicolumn{2}{|c|}{fchownat} \\
    \multicolumn{2}{|c|}{lchown} \\
    \multicolumn{2}{|c|}{ptrace} \\
    \multicolumn{2}{|c|}{setuid} \\
    \multicolumn{2}{|c|}{setgid} \\
    \multicolumn{2}{|c|}{setreuid} \\
    \multicolumn{2}{|c|}{setregid} \\
    \multicolumn{2}{|c|}{setfsuid} \\
    \multicolumn{2}{|c|}{setfsgid} \\
    \multicolumn{2}{|c|}{kill} \\
    \multicolumn{2}{|c|}{tkill} \\
    \multicolumn{2}{|c|}{tgkill} \\
    \multicolumn{2}{|c|}{mknod} \\
    \multicolumn{2}{|c|}{mknodat} \\
    \multicolumn{2}{|c|}{mount} \\
    \multicolumn{2}{|c|}{umount} \\
    \multicolumn{2}{|c|}{init\_module} \\
    \multicolumn{2}{|c|}{finit\_module} \\
    \multicolumn{2}{|c|}{delete\_module} \\
    \multicolumn{2}{|c|}{connect} \\
    \multicolumn{2}{|c|}{accept} \\
    \multicolumn{2}{|c|}{accept4} \\
    \multicolumn{2}{|c|}{listen} \\
    \multicolumn{2}{|c|}{process\_vm\_readv} \\
    \multicolumn{2}{|c|}{process\_vm\_writev} \\
    \multicolumn{2}{|c|}{inotify\_add\_watch} \\
    \multicolumn{2}{|c|}{memfd\_create} \\
    \hline
\end{tabular}
\end{table}

\begin{table*}[thb!]
\footnotesize
\centering
\caption{Android API functions in the multi-Layer events set.}
\label{table:3}
\begin{tabular} {|c|c|}
    \hline
    Class & Method \\
    \hline
    android.telephony.TelephonyManager & getImei \\
    android.telephony.TelephonyManager & getSubscriberId \\
    android.telephony.TelephonyManager & getLine1Number \\
    android.telephony.TelephonyManager & getNetworkOperatorName \\
    android.telephony.TelephonyManager & getNetworkCountryIso \\
    android.telephony.TelephonyManager & getCellLocation \\
    android.telephony.TelephonyManager & getAllCellInfo \\
    android.telephony.TelephonyManager & listen \\
    android.os.Debug & isDebuggerConnected \\
    android.app.SharedPreferencesImpl\$EditorImpl & putString \\
    android.app.SharedPreferencesImpl\$EditorImpl & putBoolean \\
    android.app.SharedPreferencesImpl\$EditorImpl & putInt \\
    android.app.SharedPreferencesImpl\$EditorImpl & putLong \\
    android.app.SharedPreferencesImpl\$EditorImpl & putFloat \\
    android.app.ActivityThread & handleReceiver \\
    dalvik.system.BaseDexClassLoader & findClass \\
    dalvik.system.BaseDexClassLoader & findResource \\
    dalvik.system.BaseDexClassLoader & findResources \\
    dalvik.system.BaseDexClassLoader & findLibrary \\
    java.lang.ClassLoader & loadClass \\
    android.app.ApplicationPackageManager & setComponentEnabledSetting \\
    android.app.NotificationManager & notify \\
    android.util.Base64 & decode \\
    android.util.Base64 & encode \\
    android.util.Base64 & encodeToString \\
    android.content.ContentResolver & query \\
    android.content.ContentResolver & registerContentObserver \\
    android.content.ContentResolver & insert \\
    android.accounts.AccountManager & getAccountsByType \\
    android.accounts.AccountManager & getAccounts \\
    android.location.Location & getLatitude \\
    android.location.Location & getLongitude \\
    android.content.ContentResolver & delete \\
    android.media.MediaRecorder & start \\
    android.app.ApplicationPackageManager & getInstalledPackages \\
    android.app.ActivityManager & getRunningAppProcesses \\
    android.app.ActivityManager & getRunningTasks \\
    dalvik.system.DexFile & openDexFile \\
    android.content.ContextWrapper & startService \\
    android.content.ContextWrapper & startActivity \\
    android.view.View & setOnClickListener \\
    java.lang.reflect.Method & invoke \\
    android.os.PowerManager & newWakeLock \\
    android.view.WindowManager & addView \\
    android.content.res.AssetManager & open \\
    android.content.res.AssetManager & openNonAssetFd \\
    android.app.ContextImpl & getSystemService \\
    android.app.usage.UsageStatsManager & queryUsageStats \\
    java.lang.Thread & sleep \\
    java.lang.reflect.Proxy & newProxyInstance \\
    \hline
\end{tabular}
\end{table*}

\end{document}